\documentclass[aps,prb,twocolumn,superscriptaddress,showpacs]{revtex4}

\usepackage{graphicx}% Include figure files

\begin{document}

%\preprint{}

\title{Interchain coupling effects on photoinduced phase transitions between neutral and ionic phases in an extended Hubbard model with alternating potentials and an electron-lattice coupling}

% Please use the appropriate macro for each type of information

\author{K. Yonemitsu}
%\thanks{}
\affiliation{Institute for Molecular Science, Okazaki 444-8585, Japan}
\affiliation{Department of Functional Molecular Science, Graduate School for Advanced Studies, Okazaki 444-8585, Japan}


\date{\today}

\begin{abstract}
Dynamics of ionic-to-neutral and neutral-to-ionic phase transitions induced by intrachain charge-transfer photoexcitations are studied in a quasi-one-dimensional extended Hubbard model with alternating potentials and an electron-lattice coupling for mixed-stack charge-transfer complexes. For interchain couplings, we use electron-electron interactions previously estimated for TTF-CA (TTF=tetrathiafulvalene, CA=chloranil). Photoexcitation is introduced by a pulse of oscillating electric field. The time-dependent Hartree-Fock approximation is used for the electronic part, and the classical approximation for the lattice part. 
In the ionic-to-neutral transition, the transferred charge density is a strongly nonlinear function of the photoexcitation density, which is characterized by the presence of a threshold. 
With substantial interchain couplings comparable to those in TTF-CA, the interchain correlation is strong during the transition. Neutral domains in nearby chains simultaneously grow even if their nucleation is delayed by reducing the amplitude of the electric field. With weaker interchain couplings, the growing processes are in phase only when the amplitude of the electric field is large. Thus, the experimentally observed, coherent motion of a macroscopic neutral-ionic domain boundary is allowed to emerge by such substantial interchain couplings. 
In the neutral-to-ionic transition, by contrast, the transferred charge density is almost a linear function of the photoexcitation density. 
Interchain electron-electron interactions make the function slightly nonlinear, but the uncooperative situation is almost unchanged and consistent with the experimental findings. 
\end{abstract}

%\keywords{}

\maketitle

\section{INTRODUCTION \label{sec:introduction}}

Photoinduced phase transitions have attracted much attention. \cite{nasu_book97} Initially local structural deformations trigger a macroscopic change in dielectric, optical, and/or magnetic properties. The transition dynamics are recently studied in different molecular materials, revealing the origin of cooperativity responsible for the proliferation of induced electronic states. \cite{nasu_book04} These researches would lead to developing techniques for dynamically controlling the electronic and structural properties and their coherence in such materials. 

The mixed-stack organic charge-transfer complex, TTF-CA, is one of the most intensively studied materials among them. The donor TTF and acceptor CA molecules are alternately stacked along the most conducting axis. At low temperature or under high pressure with contraction, they are ionic due to the long-range Coulomb interaction. \cite{torrance_prl81} Otherwise, they are neutral due to the difference between the ionization potential of the donor molecule and the electron affinity of the acceptor molecule. In the ionic phase at ambient pressure, these molecules are dimerized. 

Both ionic-to-neutral and neutral-to-ionic transitions are induced by photoirradiation near the energy of the localized intramolecular excited state of TTF molecules. \cite{koshihara_jpcb99} In the latter transition, the photoinduced phase is recently shown by X-ray diffraction to be accompanied with three-dimensionally ordered ferroelectric polarizations. \cite{collet_s03} The ionic-to-neutral transition induced by intrachain charge-transfer photoexcitations has a threshold in the photoexcitation density, below which a macroscopic neutral domain cannot be generated. \cite{suzuki_prb99} 

Ultrafast optical switching from the ionic to the neutral states is observed after a resonant excitation of the charge-transfer band polarized along the stacking axis. \cite{iwai_prl02} Recently, experimental results have been summarized covering both the ionic-to-neutral and the neutral-to-ionic transitions with different excitation energies, excitation densities, and temperatures. \cite{okamoto_prb04} In particular, the dynamics of the neutral-to-ionic transition induced by charge-transfer photoexcitations polarized along the stacking axis is clearly different from that of the ionic-to-neutral one. Although ionic domains are initially produced by lights, they quickly decay even if the excitation density is high. The initial conversion fraction is a linear function of the excitation density. 

The photoinduced phase transitions in TTF-CA have been studied also theoretically, \cite{huai_jpsj00,miyashita_jpsj03} and the origin of the different dynamics in the two transitions is discussed. \cite{yonemitsu_jpsj04a,yonemitsu_jpsj04b} However, these studies are limited to one-dimensional electron models. Although the short-time behavior may not be so sensitive to interchain couplings, they would eventually alter the long-time dynamics. For instance, they are crucial to the emergence of coherence, which can be manifested by introducing a double pulse. \cite{yonemitsu_jpsj04c,yonemitsu_error04} The effect of interchain elastic couplings was considered. \cite{huai_jpsj00} However, interchain electron-electron interactions are much stronger and would be much more influential. The \textit{ab initio} quantum chemical calculation for the intramolecular charge distribution in TTF and CA molecules \cite{kawamoto_prb01} suggests that the interchain electrostatic energies between neighboring molecules are smaller than but comparable to the intrachain ones. 

In this paper, we add interchain electron-electron interactions to the previous model for TTF-CA and study their effects on the phase transition dynamics induced by intrachain charge-transfer photoexcitations. The qualitative difference found in the purely one-dimensional model is basically unchanged: the ionic-to-neutral transition proceeds in a cooperative manner, whereas the neutral-to-ionic one in an uncooperative manner. However, the dynamics of the ionic-to-neutral transition strongly depend on the strength of interchain couplings. Indeed, substantially strong interchain electron-electron interactions are important to maintain the coherent motion of the neutral-ionic domain boundary. They allow neutral domains in nearby chains to grow simultaneously even if their nucleation is delayed by reducing the amplitude of the electric field. 

\section{QUASI-ONE-DIMENSIONAL MODEL FOR MIXED-STACK CT COMPLEXES \label{sec:model}}

For electrons in the highest occupied molecular orbital (HOMO) at donor sites and the lowest unoccupied molecular orbital (LUMO) at acceptor sites, we use a quasi-one-dimensional extended Hubbard model with alternating potentials and an electron-lattice coupling at half filling, \cite{yonemitsu_jpcs05}
\begin{equation}
H = \sum_{l=1}^{L} \left( \sum_{j=1}^{M}  H_{l,j} 
                                   + \sum_{j=1}^{M-1} H_{l,j,j+1} \right)
\;,
\end{equation}
with the intrachain component,
\begin{eqnarray}
H_{l,j} & = & -t_0 \sum_\sigma 
   \left( c^{\dagger}_{l,j,\sigma}c_{l+1,j,\sigma} 
        + \text{h.c.} \right) 
   + (-1)^{l} \frac{d}{2} n_{l,j} \nonumber \\ & & 
   + U n_{l,j,\uparrow} n_{l,j,\downarrow} 
   + \bar{V}_{l,j} \delta n_{l,j} \delta n_{l+1,j} \nonumber \\ & & 
   + \frac{k_1}{2}y_{l,j}^{2} + \frac{k_2}{4}y_{l,j}^{4} 
   + \frac{ m_{l,j} }{2} \dot{u}_{l,j}^{2} 
\;,
\end{eqnarray}
and with the interchain component,
\begin{eqnarray}
H_{l,j,j+1} & = &
   U_{\text{p}}  \delta n_{l,j} \delta n_{l,j+1} 
 + V_{\text{p1}} \delta n_{l,j+1} \delta n_{l+1,j} \nonumber \\ & & 
 + V_{\text{p2}} \delta n_{l,j} \delta n_{l+1,j+1} 
\;,
\end{eqnarray}
where, $ c^{\dagger}_{l,j,\sigma} $  ($ c_{l,j,\sigma} $) is the creation (annihilation) operator of an electron with spin $\sigma$ at site $l$ of chain $j$, $ n_{l,j,\sigma} = c^{\dagger}_{l,j,\sigma} c_{l,j,\sigma} $, $ n_{l,j} = n_{l,j,\uparrow} + n_{l,j,\downarrow} $, $ u_{l,j} $ is the dimensionless lattice displacement of the molecule from its equidistant position along the chain, and $ y_{l,j} = u_{l+1,j} - u_{l,j} $. The distance between the $ l $th and ($ l $+1)th molecules is then given by $ r_{l,j} = r_0 ( 1 + u_{l+1,j} - u_{l,j} ) $ along the $ j $th chain, where $ r_0 $ is the averaged distance between the neighboring molecules. The intersite Coulomb interactions are not between the electron densities, $ n_{l,j} $, but between the excess electron densities, $ \delta n_{l,j} $, which we define as $ \delta n_{l,j} = n_{l,j} - 2 $ for odd $ l $ (at a donor site) and $ \delta n_{l,j} = n_{l,j} $ for even $ l $ (at an acceptor site). The total charge of the donor molecule at site $ l $ is then given by $ -e \delta n_{l,j} = +e(2- n_{l,j}) $, while that of the acceptor molecule by $ -e \delta n_{l,j} = -en_{l,j} $. 

The nearest-neighbor interaction strength along the chain $ \bar{V}_{l,j} $ depends on the bond length $ y_{l,j} $, $ \bar{V}_{l,j} = V + \beta_{2} y^{2}_{l,j} $, where $ V $ is for the regular lattice. 
The parameter $ t_0 $ denotes the nearest-neighbor transfer integral, $ d $ the level difference between the neighboring orbitals in the neutral limit, and $ U $ the on-site repulsion strength. 
The elastic energy is expanded up to the fourth order with the parameters $ k_{1} $ and $ k_{2} $. 
The quantity $ m_{l,j} $ denotes the molecular mass. 
The interchain interactions are characterized by the repulsion strength $ U_{\text{p}} $ for neighboring donor molecules and for neighboring acceptor molecules, $ V_{\text{p1}} $ and $ V_{\text{p2}} $ for neighboring donor and acceptor molecules. Because of the molecular tilting, $ V_{\text{p1}} $ and $ V_{\text{p2}} $ are not equal. 
The number of sites in a periodic chain is denoted by $ L $, and that of chains by $ M $. 

\begin{figure}
\includegraphics[height=4cm]{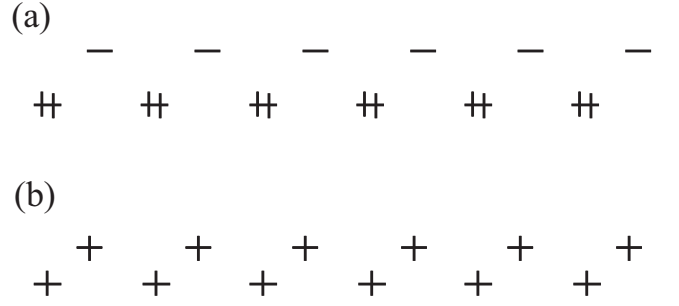}
\caption{Schematic electronic and lattice structures in (a) neutral and (b) ionic phases.
\label{fig:NI}}
\end{figure}

In the neutral phase, the orbital of the donor molecule is nearly doubly occupied, while that of the acceptor molecule is nearly empty [Fig.~\ref{fig:NI}(a)]. The total charge of any molecule is close to zero. In the ionic phase, both orbitals are nearly singly occupied [Fig.~\ref{fig:NI}(b)]. The total charges of the donor and acceptor molecules are close to $\pm e$.
%% Fig.01
The ionicity is defined as $ \rho = 1 + 1/(LM) \sum_{l=1}^{L} \sum_{j=1}^{M} (-1)^{l} \langle n_{l,j} \rangle $. 

\section{EFFECTS OF INTERCHAIN COUPLINGS \label{sec:interchain}}

In this section only, the effects of the interchain couplings are discussed for $ t_0 $=$ \beta_2 $=0. The edge chains $ j $=1 and $ j $=$ M $ are ignored unless otherwise stated. The total energy per donor-acceptor unit is $ E_{\text{N}} = -d + U $ in the neutral phase [Fig.~\ref{fig:NI}(a)], and $ E_{\text{I}} = -2V +2U_{\text{p}} -2V_{\text{p1}} -2V_{\text{p2}} $ in the ionic phase [Fig.~\ref{fig:NI}(b)]. The energy difference is then given by 
\begin{equation}
E_{\text{N}} - E_{\text{I}} = U + 2V - d_{\text{ren}} 
\;,
\end{equation}
with 
\begin{equation}
d_{\text{ren}} = d + 2U_{\text{p}} -2V_{\text{p1}} -2V_{\text{p2}} 
\;.
\end{equation}

The \textit{ab initio} quantum chemical calculation has estimated the electrostatic energies between neighboring molecules as $ V $=1.26 eV, $ U_{\text{p}} $=0.94 eV, and $ V_{\text{p1}} $=1.08 eV for TTF-CA. \cite{kawamoto_prb01} The inequality $ V_{\text{p1}} > U_{\text{p}} $, i.e., the dominance of the interchain attraction over the interchain repulsion in the ionic phase, is responsible \cite{kishine_prb04} for the phase diagram containing the paraelectric ionic phase \cite{lemee_prl97} and for the discontinuous change of the lattice parameter $ b $ at the neutral-ionic transition. \cite{luty_epl02} The inequality is due to the molecular tilting, so that $ V_{\text{p2}} $ is much smaller than $ V_{\text{p1}} $. Because the above estimations are based on  the intramolecular charge distribution for isolated molecules, the estimated values should be regarded as upper bounds for the interaction strengths. In numerical calculations, we will fix $ U_{\text{p}}/V_{\text{p1}} $=0.9 and $ V_{\text{p2}} $=0 and vary $ U_{\text{p}} $ and $ V_{\text{p1}} $. Here, we do not fix them for general discussions. 

\begin{figure}
\includegraphics[height=3cm]{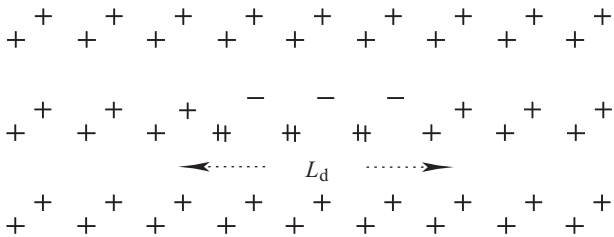}
\caption{Neutral domain in ionic background.
\label{fig:N_in_I}}
\end{figure}

To create a neutral domain of $ L_{\text{d}} $ donor-acceptor units in the ionic background (Fig.~\ref{fig:N_in_I}), it costs 
\begin{equation}
V + L_{\text{d}} ( 2 V_{\text{p1}} + 2 V_{\text{p2}} - 2 U_{\text{p}} ) 
\;, \label{eq:N_in_I}
\end{equation}
at the phase boundary, $ E_{\text{N}} $=$ E_{\text{I}} $. 
%% Fig.02
The domain wall energy is then given by $ E_{\text{DW}} = V/2 $, \cite{yonemitsu_error05} as in the one-dimensional case. \cite{nagaosa_jpsj86b} The confinement of a metastable domain is given by the $ L_{\text{d}} $-dependent term, in which the large contribution from the $ U_{\text{p}} $ term nearly cancels out the largest one from the $ V_{\text{p1}} $ term due to the relation, $ V_{\text{p1}} \agt U_{\text{p}} \gg V_{\text{p2}} $. The fact that they are comparable is important to facilitate the growth of metastable domains. 

\begin{figure}
\includegraphics[height=7cm]{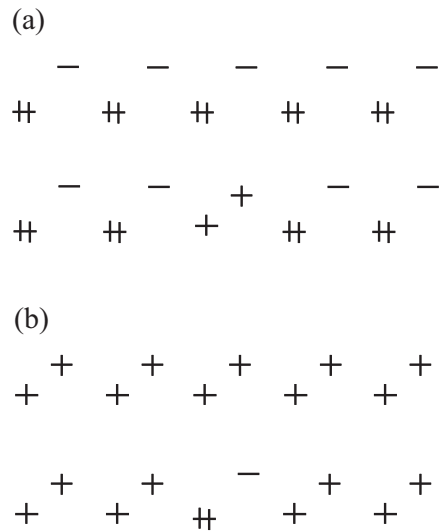}
\caption{Intrachain charge-transfer excitations in (a) neutral and (b) ionic phases.
\label{fig:opt_NI}}
\end{figure}

The optical gap for the excitation along the stacking axis is given by 
\begin{equation}
E^{\text{opt}}_{\text{N}} 
= d_{\text{ren}} -U -V -2U_{\text{p}} +2V_{\text{p1}} +2V_{\text{p2}} 
\;, \label{eq:E_opt_N}
\end{equation}
in the neutral phase [Fig.~\ref{fig:opt_NI}(a)], and by 
\begin{equation}
E^{\text{opt}}_{\text{I}} 
= -d_{\text{ren}} +U +3V -2U_{\text{p}} +2V_{\text{p1}} +2V_{\text{p2}} 
\;, \label{eq:E_opt_I}
\end{equation}
in the ionic phase [Fig.~\ref{fig:opt_NI}(b)].
%% Fig.03
Here also, the optical gaps are not drastically modified by the interchain couplings owing to the relation, $ V_{\text{p1}} \agt U_{\text{p}} \gg V_{\text{p2}} $. For $ d_{\text{ren}} $, we will employ the value that was previously used for $ d $ in the one-dimensional case. Otherwise, we do not need to modify the model parameters estimated previously. Now, we consider the edge chains $ j $=1 and $ j $=$ M $. The above equation for the optical gap is not modified in the neutral phase. In the ionic phase, the optical gap at the edge chains is given by 
\begin{equation}
E^{\text{opt}}_{\text{I}} 
= -d_{\text{ren}} +U +3V 
\;, \label{eq:E_opt_I_edge}
\end{equation}
which is slightly smaller than the bulk contribution. 

\section{METHOD OF CALCULATING DYNAMICS \label{sec:method}}

The method is the same as that used previously, \cite{yonemitsu_jpsj04a,yonemitsu_jpsj04b,yonemitsu_jpsj04c} including the time-dependent Schr\"odinger equation for the electronic part within the unrestricted HF approximation, the classical equation of motion for the lattice part according to the Hellmann-Feynman theorem, the addition of random numbers to the initial $ y_{l,j} $ and $ \dot{u}_{l,j} $ values obeying the Boltzmann distribution at a fictitious temperature $ T $, and the incorporation of a time-dependent electric field, 
\begin{equation}
E(t) = E_{\text{ext}} \sin \omega_{\text{ext}} t
\;,
\end{equation}
with amplitude $ E_{\text{ext}} $ and frequency $ \omega_{\text{ext}} $ for $ 0 < t < 2\pi N_{\text{ext}}/\omega_{\text{ext}} $ with integer $ N_{\text{ext}} $, into the Peierls phase of the transfer integral. $ E (t) $ is zero otherwise.

Because the infinitesimal deviations from the static self-consistent HF solution are equivalent to individual/collective excitations coupled with phonons in the random phase approximation (RPA), \cite{yonemitsu_prb93} the present time-dependent HF calculations treating finite deviations naturally go beyond the RPA including finite lifetimes of excitations. \cite{yonemitsu_jpsj04a} The linear absorption spectra have peaks at the energies of excitons in the RPA, which are broadened owing to interactions among excitons and other excitations. The RPA itself is beyond the Tamm-Dancoff approximation, which is nothing but the single-excitation configuration-interaction method. In the previous section, the effects of interchain couplings are discussed in the strong-coupling limit. They are described in the Tamm-Dancoff approximation also. Indeed, the present HF method produces the linear absorption spectra that are consistent with these strong-coupling estimations: it is not really a weak-coupling approach but a small-amplitude-fluctuation approach. It goes beyond the rigid-band picture by containing dynamical fluctuations but cannot qualitatively describe quantum spin fluctuations after artificial breaking of the spin rotational symmetry. Because the relevant energy scale is large in the photoinduced processes, the poor approximation for spin fluctuations would not qualitatively alter the charge-lattice coupled dynamics of present concern.

In the one-dimensional case, the model parameters are chosen to reproduce the {\it ab initio} estimation of the transfer integral and the measured values of the ionicity, the dimerization, and the absorption spectra. \cite{huai_jpsj00} We employ basically the same parameters as before \cite{miyashita_jpsj03,yonemitsu_jpsj04a,yonemitsu_jpsj04b,yonemitsu_jpsj04c} for $ M $=10 chains of $ L $=100 sites: $ t_0 $=0.17 eV, $ U $=1.528 eV, $ V $=0.604 eV (when the ionic phase is photoexcited) or $ V $=0.600 eV (when the neutral phase is photoexcited), $ d_{\text{ren}} $=2.716 eV  $ \beta_2 $=8.54 eV, $ k_1 $=4.86 eV, $ k_2 $=3400 eV, and the bare phonon energy $ \omega_{\text{opt}}  \equiv (1/r_0)(2 k_1/m_r)^{1/2} $=0.0192 eV \cite{miyashita_jpsj03}. Here, the reduced mass $ m_r $ is defined as $ m_r = m_{\text{D}} m_{\text{A}} / ( m_{\text{D}} + m_{\text{A}} ) $ with $ m_{\text{D}} $ for the donor molecule and $ m_{\text{A}} $ for the acceptor molecule. Note that the phonon energy used here is a couple of times larger than that in TTF-CA. The main results are not altered by the choice of its value. 

If we adopt the ratios $ U_{\text{p}}/V $ and $ V_{\text{p1}}/V $ evaluated by the \textit{ab initio} quantum chemical calculation \cite{kawamoto_prb01} and the value $ V $=0.604 eV above for the ionic phase, the interchain couplings are roughly estimated as $ U_{\text{p}} $=0.45 eV and $ V_{\text{p1}} $=0.50 eV. Hereafter these parameters are written in units of eV. The ionicity in the static self-consistent HF solution is 0.20 (0.18) for $ V_{\text{p1}} $=0 ($ V_{\text{p1}} $=0.50) in the neutral phase and 0.95 (0.97) for $ V_{\text{p1}} $=0 ($ V_{\text{p1}} $=0.50) in the ionic phase, which are closer to the strong-coupling limits than the experimentally estimated values of about 0.3 and 0.7. 

\section{RESULTS \label{sec:results}}

\subsection{Ionic-to-neutral transition \label{subsec:I-to-N}}

The real-time dynamics shown later depend on the frequency of the electric field. So, the linear absorption spectra in the ionic phase are calculated for reference. Without interchain couplings, the position of the absorption peak is located at $ \omega_{\text{ext}} \sim 32 \omega_{\text{opt}} \sim $0.61 eV. \cite{yonemitsu_jpsj04a} With interchain couplings $ U_{\text{p}} $=0.45 eV, $ V_{\text{p1}} $=0.5 eV and $ V_{\text{p2}} $=0, it would be shifted upward by $ ( 2 V_{\text{p1}} + 2 V_{\text{p2}} - 2 U_{\text{p}} ) $=0.1 eV$ \sim 5 \omega_{\text{opt}} $ according to Eq.~(\ref{eq:E_opt_I}) in the strong-coupling limit. The highest peak is actually located at $  \omega_{\text{ext}} \sim 38 \omega_{\text{opt}} $, corresponding to the charge-transfer excitons in the bulk. 
The second highest peak is at $ \omega_{\text{ext}} \sim 32 \omega_{\text{opt}} $ owing to the charge-transfer excitons at the edges. The latter position is the same with the one-dimensional case, as Eq.~(\ref{eq:E_opt_I_edge}) suggests. These assignments are confirmed by the fact that resonant photoexcitations produce neutral domains in the respective positions. 

When the ionic phase is photoexcited, the ionicity $ \rho_{l,j} $, defined as $ \rho_{l,j} = 1 + (-1)^l ( - \langle n_{l-1,j} \rangle + 2 \langle n_{l,j} \rangle - \langle n_{l+1,j} \rangle )/4 $, changes abruptly from a large value to a small value during the transition. \cite{yonemitsu_jpsj04a} Of course the timing of this change depends on the place. Thus, it is easy to trace the development of neutral domains. After a neutral domain is created in a chain, it may grow to convert the whole chain into the neutral phase or may be suppressed by nearby ionic chains to return to the ionic phase. For each chain, the final state is either neutral or ionic. However, it is possible for the final state to be a mixture of neutral and ionic chains if the supplied energy from the oscillating electric field is insufficient. Below we plot the number of chains converted from the ionic phase to the neutral phase among ten chains. Once a chain is converted, further application of the electric field may increase the ionicity even when the transition is not completed in other chains. This is not really a photoinduced effect, but rather a thermal effect in that the excess energy randomizes the electron distribution. \cite{yonemitsu_jpsj04a} It indeed happens for isolated or weakly coupled chains. Therefore, we have carefully looked into the evolution of ionicity at each chain. 

\begin{figure}
\includegraphics[height=6cm]{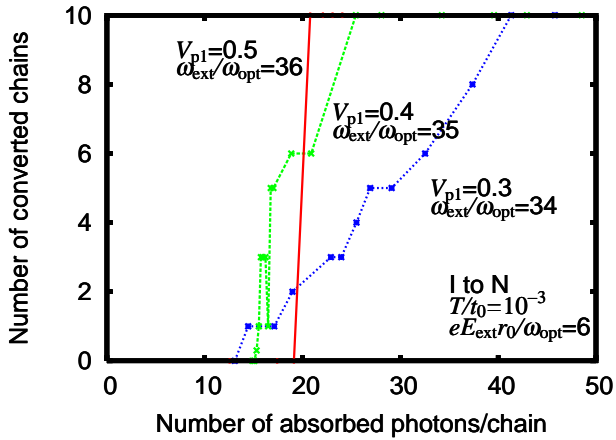}
\caption{(Color online) Number of converted chains, as a function of number of absorbed photons per chain, for different $ V_{\text{p1}} $ and $ \omega_{\text{ext}}/\omega_{\text{opt}} $. The electric field with $ eE_{\text{ext}}r_0/\omega_{\text{opt}} $=6 is applied to the ionic phase at $ T/t_0 $=10$^{-3}$. 
\label{fig:Vp_dep_IN}}
\end{figure}

The number of photoconverted chains is plotted in Fig.~\ref{fig:Vp_dep_IN} as a function of the number of absorbed photons per chain, which is obtained by dividing the increment of the total energy per 100-site chain by the frequency of the electric field. 
%% Fig.04
The frequency of the electric field is so varied with interchain couplings, $ V_{\text{p1}} $=0.3, 0.4, and 0.5 with $ U_{\text{p}}/V_{\text{p1}} $=0.9, that the energy of photons relative to that of the linear absorption peak is almost unchanged. To convert only one chain among ten, the strongly coupled chains need the higher density of photons. This is reasonable because the strongly coupled chains need a higher energy to create a neutral domain in the ionic background, according to Eq.~(\ref{eq:N_in_I}). To convert the whole system, however, the strongly coupled chains need a much lower density of photons than the weakly coupled chains. Once a neutral domain is nucleated, other neutral domains are almost simultaneously created in the neighboring chains to grow spontaneously. Their growth cannot be stopped even if the electric field is switched off immediately after the appearance of the first domain. As a consequence, the final state in the strong-coupling case is either a globally neutral one or a globally ionic one. That is why the corresponding plot is a step function. 

\begin{figure}
\includegraphics[height=6cm]{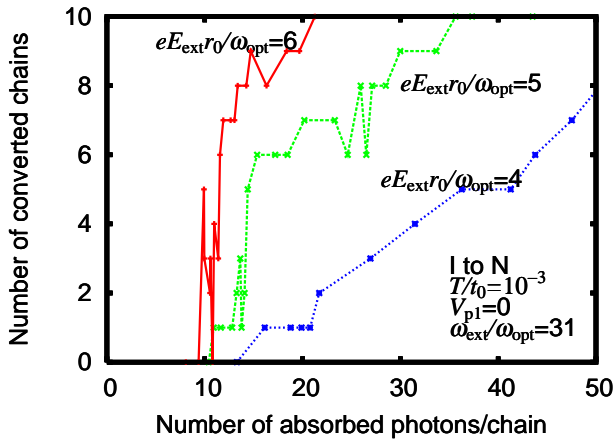}
\caption{(Color online) Number of converted chains, as a function of number of absorbed photons per chain, for isolated chains, $ V_{\text{p1}} $=0, with $ \omega_{\text{ext}}/\omega_{\text{opt}} $=31. The electric fields with $ eE_{\text{ext}}r_0/\omega_{\text{opt}} $=4, 5, and 6 are applied to the ionic phase at $ T/t_0 $=10$^{-3}$. 
\label{fig:Vp_0_IN}}
\end{figure}

In order to see how the mixture of neutral and ionic chains are produced for weak interchain interactions, we uncouple the chains by setting $ V_{\text{p1}} $=0 and use the same random numbers for the initial $ y_{l,j} $ and $ \dot{u}_{l,j} $ values as in the cases of finite interchain interactions. The number of photoconverted chains is plotted in Fig.~\ref{fig:Vp_0_IN}.
%% Fig.05
Because all the chains have the same condition except for the weakly randomized, initial lattice variables, the mixture of neutral and ionic chains is due to the distribution of the threshold photoexcitation density. Photoexcitation densities in some chains are above the threshold, while those in other chains are below the threshold. The distribution is substantially broad for small amplitude of the electric field. Within a chain, the transition proceeds cooperatively once a sufficiently long neutral domain is nucleated. The nucleation processes are randomly triggered by lattice fluctuations, so that they occur at different times. Different chains have absorbed different density of photons before the nucleation process. The distribution of the threshold photoexcitation density becomes narrower as the amplitude of the electric field increases. In Fig.~\ref{fig:Vp_0_IN}, some plots are not monotonic, where an initially created neutral domain happens to be completely annihilated by continuing photoirradiation and another neutral domain is created later in the same chain but at a different place, which eventually grows and finally covers the whole chain. \cite{yonemitsu_jpsj04a} 

For weak interchain electron-electron interactions, the relation between the number of photoconverted chains and the number of absorbed photons is similar to that for isolated chains (not shown). Indeed, for small amplitude of the electric field, the range of the photoexcitation density for producing a mixture of neutral and ionic chains is wide. As the amplitude of the electric field increases, this range becomes narrower. The interchain correlation is very weak during the transition as long as the electric field is not so strong. However, the way in which the transition proceeds is different from that for isolated chains. When the frequency of the electric field is closer to the energy of charge-transfer excitons in the bulk than that at the edges, a first neutral domain is created around the central chains by the low density of photons. If the electric field is switched off immediately after the first domain appears, only the chain containing this domain is converted into a neutral one, and the residual chains remain ionic. By continuing the application of the electric field, next domains are created in the neighboring chains. The positions of their appearances are random along the chains for weak interchain electron-electron interactions. 

Comparing the results for $ V_{\text{p1}} $=0.3 and those for $ V_{\text{p1}} $=0.4, \cite{yonemitsu_jpcs05} we find that the interchain electron-electron interactions increase both the probability of suppression after the appearance of a first domain by neighboring ionic chains and the probability of nucleation of a second domain near the first one in the neighboring chains. The interchain correlation is accordingly increased during the transition dynamics. Nevertheless, the interchain correlation is generally very weak unless the amplitude of the electric field is large. For very large amplitude of the electric field, metastable domains appear quickly and almost simultaneously after the photoirradiation starts even if the interchain electron-electron interactions are weak. Such coherence that is forced by the intense electric field survives shortly and decays rather quickly. This situation may be related to the coherent oscillation observed in the photoexcited quarter-filled-band charge-ordered organic (EDO-TTF)$_2$PF$_6$ salt. \cite{chollet_s05} 

\begin{figure}
\includegraphics[height=6cm]{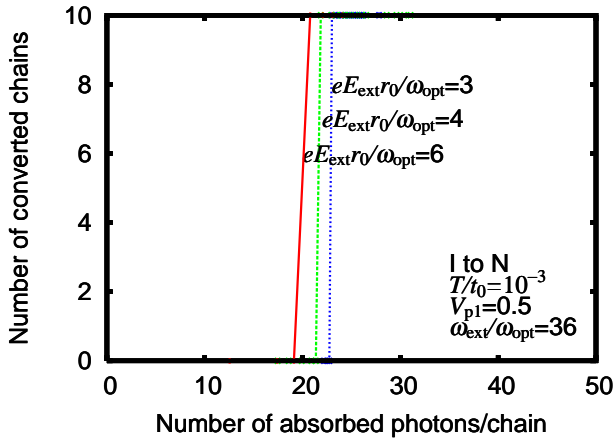}
\caption{(Color online) Number of converted chains, as a function of number of absorbed photons per chain, for strongly coupled chains, $ V_{\text{p1}} $=0.5, with $ \omega_{\text{ext}}/\omega_{\text{opt}} $=36. The electric fields with $ eE_{\text{ext}}r_0/\omega_{\text{opt}} $=3, 4, and 6 are applied to the ionic phase at $ T/t_0 $=10$^{-3}$. 
\label{fig:Vp_5_IN}}
\end{figure}

For strong interchain electron-electron interactions comparable to those in TTF-CA, the number of photoconverted chains is always a step function of the number of absorbed photons (Fig.~\ref{fig:Vp_5_IN}). 
%% Fig.06
The threshold increment of the total energy slightly increases with decreasing amplitude of the electric field. For small amplitude of the field, it takes a very long time to absorb the sufficient amount of energy to nucleate a neutral domain. Even if the appearance of a first neutral domain is delayed by such reduction of the field amplitude, it is promptly followed by the appearances of other neutral domains near the first one in the neighboring chains. It is like an avalanche phenomenon. The growth of these domains is not suppressed by the surroundings any more. The growth of neutral domains coherently proceeds, which is consistent with the experimentally observed, coherent motion of the macroscopic neutral-ionic domain boundary on a large time scale. \cite{iwai_prl02}

\subsection{Neutral-to-ionic transition \label{subsec:N-to-I}}

The linear absorption spectra in the neutral phase are first calculated before discussing the photoinduced dynamics from the neutral phase. Without interchain couplings, the position of the absorption peak is located at $ \omega_{\text{ext}} \sim 30 \omega_{\text{opt}} \sim $0.58 eV. \cite{yonemitsu_jpsj04b} With interchain couplings $ U_{\text{p}} $=0.45 eV, $ V_{\text{p1}} $=0.5 eV and $ V_{\text{p2}} $=0, it would be shifted upward again by $ ( 2 V_{\text{p1}} + 2 V_{\text{p2}} - 2 U_{\text{p}} ) $=0.1 eV$ \sim 5 \omega_{\text{opt}} $ according to Eq.~(\ref{eq:E_opt_N}) in the strong-coupling limit. The peak is actually located at $  \omega_{\text{ext}} \sim 34 \omega_{\text{opt}} $, corresponding to the charge-transfer excitons both in the bulk and at the edges. 

\begin{figure}
\includegraphics[height=6cm]{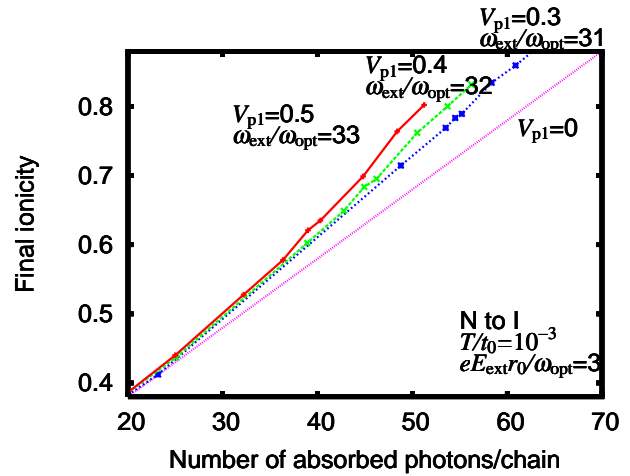}
\caption{(Color online) Final ionicity, as a function of number of absorbed photons per chain, for different $ V_{\text{p1}} $ and $ \omega_{\text{ext}}/\omega_{\text{opt}} $. The electric field with $ eE_{\text{ext}}r_0/\omega_{\text{opt}} $=3 is applied to the neutral phase at $ T/t_0 $=10$^{-3}$. 
The straight line is for isolated chains. \cite{yonemitsu_jpsj04b}
\label{fig:Vp_dep_NI}}
\end{figure}

Then, the neutral phase is photoexcited. In contrast to the ionic-to-neutral transition, the ionicity changes gradually from a small value to a large value during the neutral-to-ionic transition. After the electric field is switched off, photoinduced ionic domains spatially fluctuate \cite{yonemitsu_jpcs05} and the averaged ionicity slightly relaxes, but no large variation is observed in charge and lattice dynamics even if the transition is not completed. In other words, the photoinduced ionic domains do not grow without energy supply from the oscillating electric field. It implies that the neutral-to-ionic transition proceeds in an uncooperative manner. We plot the ionicity in the steady state in Fig.~\ref{fig:Vp_dep_NI} as a function of the number of absorbed photons per chain. 
%% Fig.07
The frequency of the electric field is so varied again with interchain couplings, $ V_{\text{p1}} $=0.3, 0.4, and 0.5 with $ U_{\text{p}}/V_{\text{p1}} $=0.9, that the energy of photons relative to that of the linear absorption peak is almost unchanged. The straight line shows the result previously obtained for isolated chains irrespective of the frequency, the amplitude, and the duration of the pulse. \cite{yonemitsu_jpsj04b} For finite interchain couplings, the final ionicity is again almost a linear function. With increasing interchain couplings, the final ionicity slightly deviates upward from the linear relation for large absorptions. It implies that they slightly enhance the cooperativity. Even for the largest couplings shown in the figure, however, the photoinduced ionic domains do not grow without energy supply. 

\section{Summary \label{sec:summary}}

Effects of interchain electron-electron interactions are studied on dynamics and nonlinear properties of phase transitions that are induced by intrachain charge-transfer photoexcitations, in a quasi-one-dimensional extended Hubbard model with alternating potentials and the electron-lattice coupling for mixed-stack charge-transfer complexes. Within the self-consistent mean-field approximation, we solve the time-dependent Schr\"odinger equation for the electronic part and the classical equation of motion for the lattice part. 

In the ionic-to-neutral transition, the conversion fraction is a strongly nonlinear function of the photoexcitation density for any interchain couplings. This cooperative property is maintained even if the electron-lattice interaction is turned off, \cite{yonemitsu_jpsj04b} so that it is presumably due to electron-electron interactions that cause the ionic state basically a Mott insulator. All the electrons are so correlated in the ionic phase that any one of them cannot easily make a first move below the threshold. Because energies are accumulated by photoirradiation of the ionic phase, each electron feels a force, but its motion is suppressed by electron-electron interactions. Once a metastable neutral domain is created above the threshold, it spontaneously grows leading to cooperative charge transfer. A similar collective charge-transport phenomenon is observed in field-effect transistors fabricated on organic single crystals of a quasi-one-dimensional Mott insulator, \cite{hasegawa_prb04} whose ambipolar characteristics are theoretically shown to be caused by balancing the correlation effect in the bulk with the Schottky barrier effect at interfaces. \cite{yonemitsu_jpsj05} 

For weak interchain electron-electron interactions, there is a range of the photoexcitation density that produces a mixture of neutral and ionic chains. This range is wide for small amplitude of the oscillating electric field because the lattice-fluctuation-induced distribution of the threshold photoexcitation density is broad. For very large amplitude of the electric field, however, metastable domains are forced to grow simultaneously. Such coherence by the intense electric field survives only shortly. 

For strong interchain electron-electron interactions comparable to those in TTF-CA, the interchain correlation is strong during the transition. Neutral domains in nearby chains simultaneously grow even if their nucleation is delayed by reducing the amplitude of the electric field. Thus, the experimentally observed, coherent motion of the macroscopic neutral-ionic domain boundary on a large time scale \cite{iwai_prl02} is a consequence of the strong interchain couplings. 

In the neutral-to-ionic transition, the conversion fraction is almost a linear function of the photoexcitation density. This property is maintained even when we preliminary added interchain elastic couplings to the present model: the energy scale of elastic interactions is so small compared with that of electron-electron interactions that elastic couplings would not largely modify the charge dynamics at least on a short time scale. Such an uncooperative property is caused by uncorrelated electrons in the neutral phase, which move individually. The supplied energy is merely consumed to transfer electrons almost independently. The growth of metastable ionic domains is not spontaneous but always forced by the external field. Interchain electron-electron interactions make the function slightly nonlinear, but the above situation is almost unchanged. This qualitative difference between the ionic-to-neutral and the neutral-to-ionic transitions is consistent with the experimental findings. \cite{okamoto_prb04}

%\appendix
%\section{}

\begin{acknowledgments}
The author is grateful to H. Cailleau, S. Iwai, S. Koshihara and H. Okamoto for showing their data prior to publication and for enlightening discussions. 
This work was supported by Grants-in-Aid for Scientific Research (C) (No. 15540354), for Scientific Research on Priority Area ``Molecular Conductors'' (No. 15073224), for Creative Scientific Research (No. 15GS0216), and NAREGI Nanoscience Project from the Ministry of Education, Culture, Sports, Science and Technology, Japan.
\end{acknowledgments}

\bibliography{pipt05a}

\begin{thebibliography}{25}
\bibitem[{yon({\natexlab{a}})}]{yonemitsu_error04}
\bibinfo{note}{In the figures of
  Refs.~\onlinecite{yonemitsu_jpsj04a,yonemitsu_jpsj04b,yonemitsu_jpsj04c}, the
  parameter $ a $ should read the averaged distance $ r_0 $, $ T/t $ the degree
  of initial lattice fluctuations $ T/t_0 $, and $ y_{\text{st}} $ the
  staggered lattice displacement $ u_{\text{st}} $.}
\bibitem[{yon({\natexlab{b}})}]{yonemitsu_error05}
\bibinfo{note}{The formula for the domain wall energy $ E_{\text{DW}} $ was
  wrong in Ref.~\onlinecite{yonemitsu_jpcs05}}.
\end{thebibliography}

\end{document}